# *Optical Spectra of Graded Nanostructured TiO$_2$ Chiral Thin Films*

F. Babaei[1,*], A. Esfandiar[2], H. Savaloni[2]

1) Department of Physics, University of Qom, Qom, Iran.
2) Department of Physics, University of Tehran, North-Kargar Street, Tehran, Iran.
*) Corresponding author: Tel: +98 251 2853311; Fax: +98 251 2854972; Email: fbabaei@qom.ac.ir

PACS: 42.70.-a; 78.20.-e; 61.05.-a; 68.37.-d

## Abstract

The rigorous coupled wave analysis method (RCWA) is applied to the graded chiral TiO$_2$ thin films in axial and non-axial excited states using the Bruggeman homogenization formalism. The dependence of the filtering frequency and polarization selectivity as a function of different structural parameters, are investigated. A good agreement with experimental results reported by other researchers is achieved.

*Keywords: graded chiral thin films; glancing angle deposition; circular selectivity*

## 1. Introduction

Chiral thin films or sculptured chiral thin films (STFs) are columnar (range betwe en 1 to 100 nm) thin films deposited on substrate with controlled azimuthal rotation, $\phi$, and tilt angle, $\alpha$, by a method known as glancing angle deposition (GLAD). In this method, the columnar structure of thin film can be designed in nano and micro scale 3D for use in optical, chemical, mechanical, biological, electrical and magnetic applications [1]. The GLAD method can be used to produce 3D photonic crystals that depending on their nano-structural dimensions they prohibit the propagation of a certain bands of electromagnetic frequency through macroscopic (Bragg) and microscopic scattering resonances [2].

The most important optical feature of chiral thin films is the circular Bragg phenomenon that unlike ordinary thin films occurs in chiral thin films. This phenomenon is studied extensively both theoretically [3-11] and experimentally [12-15]. If the handedness of the chiral thin film is the same as that of the polarization of the incident circular plane wave, then nearly all the incident light will be reflected in a narrow region of the wavelength (this is called the Bragg regime), otherwise nearly all the incident light will be transmitted. In these films (chiral or normal chiral thin films) the helical axis or the axis of non-homogeneity is normal to the substrate surface (z-axis). These films can be used to discriminate the right handed circular polarization (RCP) from the left-handed circular polarization (LCP). There is another group of chiral thin films called slanted chiral thin films in which the helical axis is tilted with respect to the normal to the substrate surface. In this kind of thin films due to the periodicity of structural parameters in both normal and parallel directions to the substrate surface, the interaction between circular Bragg phenomenon (specular diffraction) and non-specular diffraction (Rayliegh-Wood anomalies) occur which can be related to the diffraction gratings [16-19]. The study on the slanted chiral thin films is of importance in understanding the optical ray diffraction by sculptured thin films [20]. In addition, many applications have been proposed for slanted chiral thin films, such as optical beam splitters, couplers, nano-band and sub nano-band spectral-hole filters and biosensors [21-24].

A third group of chiral thin films have been introduced by Krause and Brett [25], so called graded chiral thin films which are the similar to the slanted chiral thin films with a gradient in their thickness in their nano-wire structural assembly (Fig. 1). The graded chiral thin films are produced using the GLAD system with a shadowing block positioned at the center of the rotating substrate. This shadowing object modulates the microscopic structure of the thin film and produces spatially graded chiral thin film with graded optical properties. In these films it is possible to distinguish both polarization and frequency selection simultaneously. This technique can be used to produce polarization responsive and tunable frequency optical filters, sources, and detectors for applications such as multichannel light wave systems as well as providing a graded scaffolding to support liquid crystals [25].

Optical devices with tunable optical functions (refractive index, absorption and extinction coefficient, optical activity and other functions) have always been of interest and desire in different areas of science and industry. A large body of the recent research in the field of thin films and nano-structures is devoted to the production of an optical device with filtering frequency and polarization selectivity properties in opto-electronic. However building of a device, that can have both of these properties simultaneously have been a difficult task. Similar films exist in the nature, such as animals with skins, feathers, wings of varying thickness and chiral structure which while shows different colors it also changes the polarization of the incident light [26-28]. Thin films with variable structural characteristics and filtering frequency at varying frequency along their length have already been produced using different techniques [29-32]. Bragg reflection from gratings is also a method for producing optical components with tuneable characteristics [33,34]. Cunningham et al. [30] also reported a guided-mode resonance filter _GMRF_ photonic crystal which can separate the reflection of specific wavelength bands across the width of the component. On the other hand, liquid crystals with helical molecular structures can distinguish the circular polarization of the incident light [35]. Production of nano-structures with sensitivity to the circular polarization of the incident light has also been achieved by means of GLAD technique in recent years. Krause and Brett [25] who inspired from the nature used a macroscopically shadowing effect in the GLAD technique to produce thin films with both frequency filtering property and sensitivity to the polarization of the light.

In this work the rigorous coupled wave analysis (RCWA) proposed and extended by Lakhtakia [20] for slanted chiral thin films is applied to graded chiral thin films using a computer code written in Mathematica 5.2. Reflection, transmission and circular polarization selectivity for graded chiral thin films are calculated and compared with the experimental results reported in [25]. A good agreement between our theoretical results and those of Krause and Brett [25] is achieved.

In Section 2 the finalized formalism of the rigorous coupled wave analysis is given. The numerical results are reported, discussed and compared with Krause and Brett's [25] experimental results in Section 3.

## 2. Theory

The rigorous coupled wave analysis (RCWA) method was first proposed by Moharam and Gaylord in 1981 [36] in order to obtain a near-exact solution to Maxwell equations for periodic diffracting structures. Wang and Lakhtakia [20] formalized the rigorous coupled-wave analysis for the slanted chiral thin films and obtained the following final algebraic equation:

$$
\begin{bmatrix} e^{-id[\underline{\underline{D}}_1]}[\underline{\underline{U}}_T] & [\underline{\underline{U}}_R] \\ [\underline{\underline{V}}_T] & e^{id[\underline{\underline{D}}_2]}[\underline{\underline{V}}_R] \end{bmatrix} \begin{bmatrix} [\underline{T}] \\ [\underline{R}] \end{bmatrix} = \begin{bmatrix} [\underline{\underline{U}}_A] \\ e^{id[\underline{\underline{D}}_2]}[\underline{\underline{V}}_A] \end{bmatrix} [\underline{A}] \qquad (1)
$$

where,

$$[\underline{A}] = \begin{bmatrix} a_L^{(n)} \\ a_R^{(n)} \end{bmatrix}, \quad [\underline{R}] = \begin{bmatrix} r_L^{(n)} \\ r_R^{(n)} \end{bmatrix}, \quad [\underline{T}] = \begin{bmatrix} t_L^{(n)} \\ t_R^{(n)} \end{bmatrix} \tag{2}$$

where, $\{a_L^{(n)}, a_R^{(n)}\}$, $\{r_L^{(n)}, r_R^{(n)}\}$, $\{t_L^{(n)}, t_R^{(n)}\}$ are the nth order harmonic complex amplitudes of the incident, and reflection and transmission of left- and right-handed circularly polarized light, respectively. The rest of parameters in Eq. 1 are described in detail in Section 2-2 of reference [20].

Once all $\{a_L^{(n)}, a_R^{(n)}\}$, $\{r_L^{(n)}, r_R^{(n)}\}$, $\{t_L^{(n)}, t_R^{(n)}\}$ are for all n orders have been determined using the RCWA, the nth order reflection and transmission coefficient can be obtained from:

$$r_{\sigma\sigma'} = \frac{r_{\sigma}^{(n)}}{a_{\sigma'}^{(0)}}, \quad t_{\sigma\sigma'} = \frac{t_{\sigma}^{(n)}}{t_{\sigma'}^{(0)}}, \quad \sigma, \sigma' = L, R \tag{3}$$

The nth order harmonic reflection and transmission are:

$$R_{\sigma\sigma'}^n = \frac{\text{Re}[k_z^{(n)}]}{\text{Re}[k_z^{(0)}]} \left| r_{\sigma\sigma'}^{(n)} \right|^2, \quad T_{\sigma\sigma'}^n = \frac{\text{Re}[k_z^{(n)}]}{\text{Re}[k_z^{(0)}]} \left| t_{\sigma\sigma'}^{(n)} \right|^2, \quad \sigma, \sigma' = L, R \tag{4}$$

The first subscript shows the polarization state of the reflected or transmitted and the second subscript indicates the polarization state of incident light and Re[ ] is the real part. $k_z^{(n)}$ is the z component of the wave vector for the nth order harmonic ($K_{\pm}^{(n)} = k_x^{(n)} u_x + k_y^{(0)} u_y \pm k_z^{(n)} u_z$):

$$\begin{cases} k_z^{(n)} = +\sqrt{k^2 n_{hs}^2 - (k_{xy}^{(n)})^2} \\ k_{xy}^{(n)} = +\sqrt{(k_x^{(n)})^2 + (k_x^{(0)})^2} \\ k_x^{(n)} = k_x^{(0)} + n k_x \\ k_x = \frac{\pi}{\Omega} \left| \sin a_T \right| \end{cases} \tag{5}$$

where, $k = \dfrac{2\pi}{\lambda}$, $\lambda$ is the wavelength of light in vacuum. The lower and the upper half-spaces are filled by a homogeneous, non-dissipative, dielectric medium of refractive index $n_{hs}$.

For $\alpha_T = 0$, all non-specular reflections and transmissions diminish (n # 0) and only specular reflection and transmission of zero order harmonic will appear:

$$R_{\sigma\sigma'}^0 = \left| \sum_{n \le N_t} r_{\sigma\sigma'}^{(n)} \right|^2, \quad T_{\sigma\sigma'}^0 = \left| \sum_{n \le N_t} t_{\sigma\sigma'}^{(n)} \right|^2, \quad \sigma, \sigma' = L, R \qquad (6)$$

In our calculations $N_t = 2$ was assumed, while higher $N_t$ were also investigated and it was found out that the results are the same as that for $N_t = 2$ up to the accuracy of $10^{-6}$.

Circular polarization selectivity of nth order harmonic for the graded chiral thin film is:

$$Selectivity^{(n)} = T_{LL}^{(n)} - T_{RR}^{(n)} \qquad (7)$$

Since all non-specular transmission spectra for axial excitation state were less than 0.01 (one percent), only the co-polarized transmission spectra $T_{LL}$ and $T_{RR}$ and circular polarization selectivity of the zero order harmonic were studied and reported in this work.

### 3. Results and discussions

It was assumed that a right-handed $TiO_2$ graded chiral sculptured thin film in its bulk state (rutile phase) with dielectric constants [37] given in Fig. 2 is formed, which occupies a space with refractive index of $n_{hs}$ (Fig. 1). d is considered as the film thickness at a certain distance from the edge of the shadowing block. In order to obtain the relative permittivity scalars $\varepsilon_{a,b,c}$, we used the Bruggeman homogenization formalism [38] with ellipsoidal form factors $\gamma_b^s = 2$, $\gamma_\tau^s = 20$, $\gamma_b^v = 2$ $\gamma_\tau^v = 20$, [39] in the relative permittivity of the refernce $\underline{\underline{\varepsilon}}_{ref}(\underline{r}) = \varepsilon_a \underline{u}_z \underline{u}_z + \varepsilon_b \underline{u}_x \underline{u}_x + \varepsilon_c \underline{u}_y \underline{u}_y$ ($\underline{u}_x$, $\underline{u}_y$ and $\underline{u}_z$ are Cartesian unit vectors). The homogenization process was carried out in 2 nanometer wavelength steps for each given frequency with respect to the refraction index of the TiO$_2$ at that frequency. In this way the dielectric dispersion function was implemented in the calculations. Hence, this method directly takes advantage of the experimental relative dielectric constant and avoids the use of simple dispersion model known as single-resonance Lorentzian model [40,41].

On the basis of Krause and Brett's experimental observation [25] the film thickness along a distance of 45 mm from the edge of the macroscopic shadowing block varies only about 1.5 $\mu$m (i.e., about 30 nm per mm distance or less than tenth of a complete pitch of chiral structures). with respect to this and considering that the light diameter incident on the sample surface in the spectrophotometer is about one millimetre one can assume that the interaction region of the incident light with the graded chiral thin film is a slanted chiral thin film with constant structural parameters in each given distance (of the order of millimetre) from the edge of the macroscopic shadowing block. In this paper the results of calculations for regions with more than five millimetre separation distance are reported. This choice of separation distance provide us with the opportunity to compare our theoretical results with Krause and Bret's experimental results [25], directly and the fact that two consecutive regions will have considerably different structural parameters.

In order to be able to compare our results with those of Krause and Brett, the structural parameters of their reported work was needed. In their work d and $\Omega$ increased and $\alpha_T$ decreased by going away from the edge of the macroscopic shadowing block. The void fraction $f_v$ and $\chi$ are constant for a given deposition angle.

In Fig. 3 the calculated results for the co-polarized transmission spectra $T_{LL}$ and $T_{RR}$, and circular polarization selectivity calculated for a right-handed graded chiral thin film in the axial excited state for a region at 50 mm distance from the edge of the shadowing block at a deposition angle of 70 degree are plotted. In Table 1 the experimental structural and deposition parameters adapted from Krause and Brett's [25] work for a graded chiral thin film deposited at 70 degree are given for different distances from the edge of the shadowing block. We used these parameters in our calculations.

**Table 1.** The structural parameters for a graded chiral TiO$_2$ film produced at a deposition angle of 70° (adapted from ref. [25])

| D (mm) | $\Omega$ (nm) $\Omega = \Omega_{Exp}^{BG}/\cos(\alpha_T)$ | $\alpha_T$ (°) | $\chi$ (°) | d (nm) |
|---|---|---|---|---|
| X=50mm | 200 | 6 | 37 | 18 $\Omega$ |
| X=35mm | 192 | 11 | 37 | 18 $\Omega$ |
| X=20mm | 186 | 15.5 | 37 | 18 $\Omega$ |
| X=15mm | 177 | 17.3 | 37 | 18 $\Omega$ |
| X=10mm | 173 | 17.5 | 37 | 18 $\Omega$ |
| X=5mm | 160 | 18 | 37 | 18 $\Omega$ |

D) is the distance from the edge of the shadowing block

The deposition angle of 70 degree was chosen on the basis of Sorge et al. [12] who reported the occurrence of the maximum of circular polarization selectivity at about this deposition angle. Out side the Bragg region, all reflections and transmissions show oscillations that correspond to the Fabry-Perot oscillations, which occur in a homogeneous medium because of interference of reflected light from two surfaces of the medium. These reflections are affected by both thickness and the average of the refractive index of the medium [42]. Since the handedness of the chiral thin film and the polarization direction of the left-handed circularly plane wave are not the same, the interaction in the Bragg regime is low. Therefore, in the Bragg regime the left-handed circularly plane wave unlike the right-handed circularly plane wave is almost completely transmitted through the medium (Fig. 3-a). The maximum of circular polarization selectivity is obtained at 694 nm which is in good agreement with the experimental result of Krause and Brett (690 nm) [25] (Fig. 3-b).

In Fig. 4, the calculated circular polarization selectivity spectra at different distances from the shadowing block, for a right-handed graded chiral thin film in axial excited state, produced at a deposition angle of 70 degree are depicted. In this figure in order to discriminate the spectra from each other, each spectrum is shifted to higher level by 5%. The Bragg and Raleigh-Wood wavelengths used for obtaining Fig.4 are given in Table 2.

**Table 2.** $\lambda_e^{Br}$ at different distances from the edge of the shadowing block.

| X(mm) | 5 | 10 | 15 | 20 | 35 | 50 |
|---|---|---|---|---|---|---|
| (nm) $\lambda_e^{Br}$ | 510 | 572 | 590 | 632 | 664 | 684 |

Fig. 4 shows that the peak of circular polarization selectivity (the location of appearance of circular Bragg phenomenon) shifts towards shorter wavelengths (blue shift) by decreasing the distance from the shadowing block, while the intensity at its maximum is also reduced. This is due to decreasing of the structural period (because of reduction in film thickness), increasing of tilt angle of chiral nano-wires ($\alpha_T$) (see Table 1) and the fact that the film becomes anisotropic. Therefore, graded chiral thin film while possesses frequency tenability is capable of changing the polarization of the incident light. Hence, it can simultaneously have filtering frequency and polarization selectivity properties of light.

In Fig. 5, the Maximum circular polarization selectivity calculated at a distance of 50 mm from the edge of the shadowing block, for a right-handed graded chiral thin film in axial excited state, produced at different deposition angles and according to the structural parameters given in Table 3 (adapted from Krause and Brett's [25] work) , are plotted.

**Table 3.** The structural parameters for a graded chiral $TiO_2$ film produced at different deposition angles, edge of shadow block is at X = 50 mm and thickness is 18 $\Omega$. (adapted from ref. [25])

| deposition angle(°) | $\Omega$ (nm) $\Omega = \Omega_{Exp}^{BG} / \cos(\alpha_T)$ | $\alpha_T$ (°) | $\chi$ (°) | $f_v$ (%) |
|---|---|---|---|---|
| 50 | 184 | o.1 | 40 | 28 |
| 55 | 188 | o.1 | 40 | 33 |
| 60 | 195 | o.1 | 39 | 39 |
| 65 | 197 | o.1 | 38 | 45 |
| 70 | 200 | 6 | 37 | 53 |
| 75 | 227 | 20 | 35 | 62 |
| 80 | 244 | 22 | 32 | 68 |
| 85 | 253 | 25.5 | 30 | 75 |

For a shadowing block of 20 mm height and the deposition angles of less than 70 degree and at a distance of 50 mm from the edge of the shadowing block, the tilt angle was assumed to be small, because at these deposition angles the shadowing length of the block is less than 50 mm. The rise angle for deposition angles greater or equal to 70 degree is in good agreement with Tait's law [43] but at lower deposition angles the existing relationships between the rise angle and deposition angle (reported for tilted columnar structure and not for helical structures) are not suitable, because at these angles the void fraction density in the film structure reduces and the structure consists of closely packed features. Differentiation of individual chiral structures becomes more difficult. Therefore, in order to obtain results at deposition angles less than 70 degrees, the rise angle was fixed at naerly 40 degrees, so that the results satisfy the experimental observations [25] (Table 3). It should be mentioned that the data given in Table 3 for the rise angle at deposition angles less than 70 degrees (i.e., 65 and 60 degrees) belong to the fitting of our calculations to Krause and Brett data in Fig. 5, as neither numerical nor SEM results were present in [25] for these angles. On the other hand, the use of available relationships, such as tangent role which is given for columnar thin films, proved to produce unacceptably large rise angles (i.e., 50 and 60 degrees). It is worthwhile to mention that in none of the theoretical works reported so far (e.g., [20]) such large rise angles are used. In other words, the use of such large rise angles will produce structures with no resemblance to chiral thin films. The void fraction in thin films at deposition angles smaller than 75 degree can be obtained using the relationship $\rho = \rho_0 \dfrac{2\cos(\alpha_d)}{1+\cos(\alpha_d)}$ ($\rho$ and $\rho_0 = 3.9\,grcm^{-3}$ are the film and bulk densities, respectively) [25] and for deposition angles greater than 75 degree we used the experimental results reported in reference [44]. The results given in Fig. 5 are consistent with those obtained experimentally by Krause and Brett (see Fig. 9 in [25]) (the maximum of the circular polarization selectivity appears at 70 degree). At deposition angles smaller than 70 degree, according to the 5th column in Table 3 the void fraction decreases and the film becomes almost isotropic. Hence, at these deposition angles due to reduction of the intensity in the circular Bragg phenomenon the circular polarization selectivity decreases too. At deposition angles greater than 70 degree, the void fraction increases, the chirality of the film diminishes and the scattering increases, resulting in lower circular polarization selectivity [45].

Bragg wavelengths calculated for different incident angles at a distance of 50 mm from the edge of the shadowing block and used for obtaining the results in Fig. 5 are given in Table 4.

**Table 4.** $\lambda_e^{Br}$ for different incidence angles at a 50 mm distances from the edge of the shadowing block.

| (º)$\theta_{inc}$ | 50 | 55 | 60 | 65 | 70 | 75 | 80 | 85 |
|---|---|---|---|---|---|---|---|---|
| (nm)$\lambda_e^{Br}$ | 794 | 778 | 762 | 726 | 684 | 648 | 628 | 572 |

In Fig. 6 (a-c), the spectra of circular polarization selectivity obtained from our calculation for a right-handed graded TiO$_2$ chiral thin film in axial and non-axial excited states at 70 degree deposition angle for three different distances from the shadowing block (X= 10, 20 and 50 mm) are depicted. It can be observed that the circular polarization selectivity for non-axial excitation state is higher than that for axial excitation at any given distance from the edge of the shadowing block. This is due to the fact that in non-axial excitation state the incident angle of the light on the film ($\theta_{inc}$) is chosen to be the same as $\alpha_T$ of nano-wires. This choice increases the intensity of the circular Bragg phenomenon which results in increased circular polarization selectivity. Of course at distances far away from the shadowing block the difference seen in Fig. 6 diminishes due to the fact that the film structure becomes isotropic and the nano-wires grow normal to the substrate surface. Hence, if higher circular polarization selectivity at distances close to the shadowing block is required, the incident light angle on the film ($\theta_{inc}$) should be chosen to be similar to $\alpha_T$ of the nano-wires at that distance. This selection of incident angle causes a blue shift in axial excited state and a red shift in non-axial excited state.

In Fig. 7, the circular reflection spectra $R_{RR}^{-2}$ and $R_{RR}^{°}$ for a right-handed graded TiO$_2$ chiral thin film in non-axial excited states produced at 70 degree deposition angle for four different distances from the shadowing block ($\alpha_T$ = 6º, 11º, 15.5º and 17.5º) and for the Bragg wavelength given in Table 2 (Bragg wavelength is the wavelength that is obtained from the calculations in the axial excitation state) are plotted. From the comparison of $R_{RR}^{-2}$ spectra in $\Psi_{inc} = 0°$ and $\Psi_{inc} = 90°$ plots it can be deduced that at $\Psi_{inc} = 90°$ the spectra are symmetric, which is due to anomaly effects (Rayleigh-Wood anomalies phenomenon) on circular Bragg phenomenon. It should be mentioned that at $\Psi_{inc} = 0°$ the intensity and location of the circular Bragg phenomenon depends on $\alpha_T$.

From the comparison of $R_{RR}^{°}$ spectra in $\Psi_{inc} = 0°$ and $\Psi_{inc} = 90°$ plots (Fig. 7) it can be deduced that the total reflection regime ($\left| \sin\theta_{inc} \right| \in [0.55, 1]$ is insensitive to $\alpha_T$ of nano-wires or the different distances from the edge of the shadowing block and hence to the filtering frequency property. In addition, it is being reported that the total reflection regime does not depend on the handedness of the film [23].

Fig. 8 shows the refractive index of a right-handed graded $TiO_2$ chiral thin film in both axial and non-axial excited states at different deposition angles and at distance of 50 mm from the edge of the shadowing block. The refractive index of the axial excited state was calculated from $\lambda^{Bragg} = n(2\Omega)\cos(\alpha_T)$ relationship and in the non-axial state the Snell law, $\sin\theta_{inc}^{c} = \dfrac{n_{ave}}{n_{hc}}$ (the on set angle of the total reflection regime in the circular reflection spectrum $R_{RR}^{-2}$), was used. The low refractive index obtained at higher deposition angles is due to the increased void fraction.

The results presented in Fig. 8 are consistent with those of Krause and Brett [25]. The theoretical results in this work show the same behavior as those of experimental work. The small deviation at deposition angles greater than 75 degrees (Figs. 5 and 8) could be due to the fact that the following phenomena are not included in our calculations: 1) the strong scattering from the surfaces of the nanostructures at these angles, 2) disappearance of helical columns with increasing film thickness and increase of their diameter, 3) existence of two different media above (vacuum) and underneath (substrate) the film, 4) more precise data (measurements) for structural parameters, 5) instrumental (experimental) uncertainties, and other parameters. The inclusion of the above phenomena in the calculations is our future task and we expect to achieve enhanced results.

## 4. Conclusions

we used the rigorous coupled-wave analysis (RCWA) method in conjunction with the Bruggeman homogenization formalism to calculate the co-polarized reflection and transmission spectra and the circular polarization selectivity for a right-handed graded $TiO_2$ chiral thin film in both axial and non-axial excitation states. The influence of structural and deposition parameters, namely, void fraction, structural period, tilt angle, deposition angle, and distance from the edge of the shadowing block on these spectra were studied. A good agreement between our theoretical calculations and the experimental results reported by Krause and Brett [25] is achieved. It is deduced that in order to observe high circular polarization selectivity at close distances to the edge of the shadowing block, the incident angle of light $\theta_{inc}$ must be equal to $\alpha_T$ of the nano-wires. This work has indicated that the RCWA can successfully be used to characterize and predict the behavior of the graded chiral thin films.

**Acknowledgements**

This work was supported by the University of Tehran and the University of Qom.

**Figure captions**

Fig. 1. Schematic of graded nanostructured chiral thin films. $\alpha_d$, $\alpha_T$, $\Omega$ and $\chi$, are deposition angle, tilt angle, half structural period and rise angle, respectively.

Fig. 2. The dielectric constant ( $\varepsilon = \varepsilon_1 + i\varepsilon_2$ , $\varepsilon_1 = n^2 - k^2$ $and$ $\varepsilon_2 = 2\,n\,k$ ) of bulk TiO$_2$ (rutile phase), showing that the imaginary part is nearly zero in the wavelength region shown.

Fig. 3. a) $T_{LL}$ and $T_{RR}$ transmission spectra. b) Difference in transmission spectra, for an axially excited right-handed graded chiral TiO$_2$ thin films at 50 mm from the edge of shadowing block. Deposition angle is 70° and $n_{hs} = 1$ .

Fig. 4. Difference in transmission spectra for an axially excited right-handed graded chiral TiO$_2$ thin films calculated at different distances from the edge of shadowing block. Deposition angle is 70° and $n_{hs} = 1$ .

Fig. 5. Maximum selectivity calculated at 50 mm from the edge of shadowing block for different deposition angles. $n_{hs} = 1$ .

Fig. 6. Polarization selectivity spectra of a right-handed graded TiO$_2$ chiral thin film in axial and non-axial excited states at 70 degree deposition angle for three different distances from the shadowing block (X= 10, 20 and 50 mm). $n_{hs} = 1$ .

Fig. 7. Reflection spectra for a right- handed graded chiral TiO$_2$ thin film. Deposition angle is 70° and $n_{hs} = 3.5$ . Dashed curve is $R_{RR}^{-2}$ and solid curve is $R_{RR}^{0}$ .

Fig. 8. The index of refraction of a right-handed graded TiO$_2$ chiral thin film in both axial ( $\lambda^{Bragg} = n(2\Omega)\cos(\alpha_T)$ ) and non-axial excited states ( $\sin\theta_{inc}^c = \dfrac{n_{ave}}{n_{hc}}$ ) at different deposition angles and at distance of 50 mm from the edge of the shadowing block.

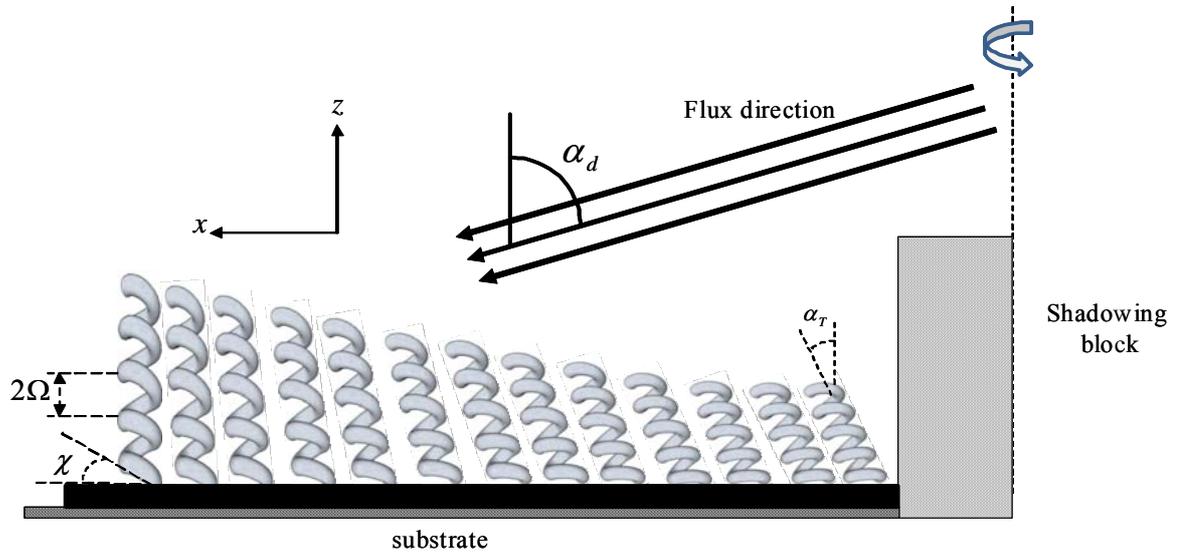

Fig. 1. Babaei et al.

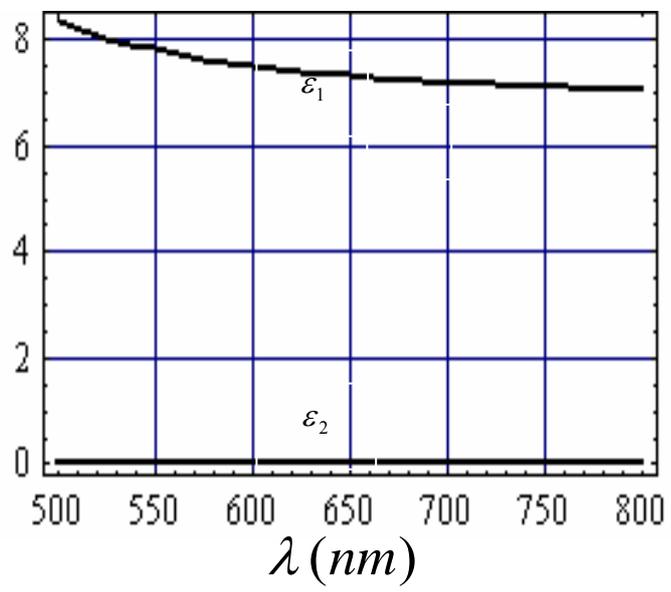

**Fig. 2. Babaei et al.**

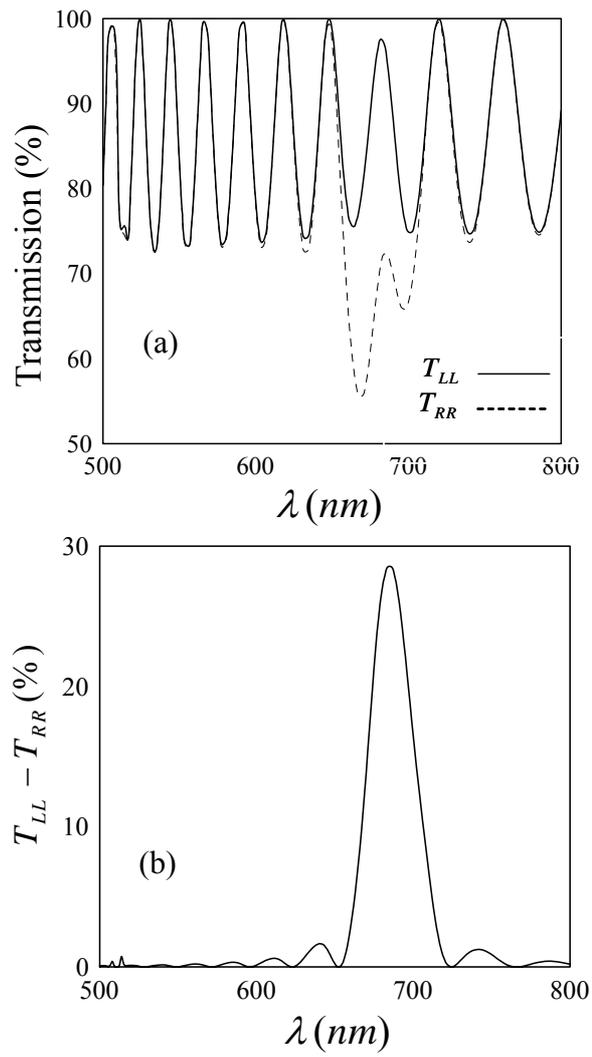

Fig. 3. Babaei et al.

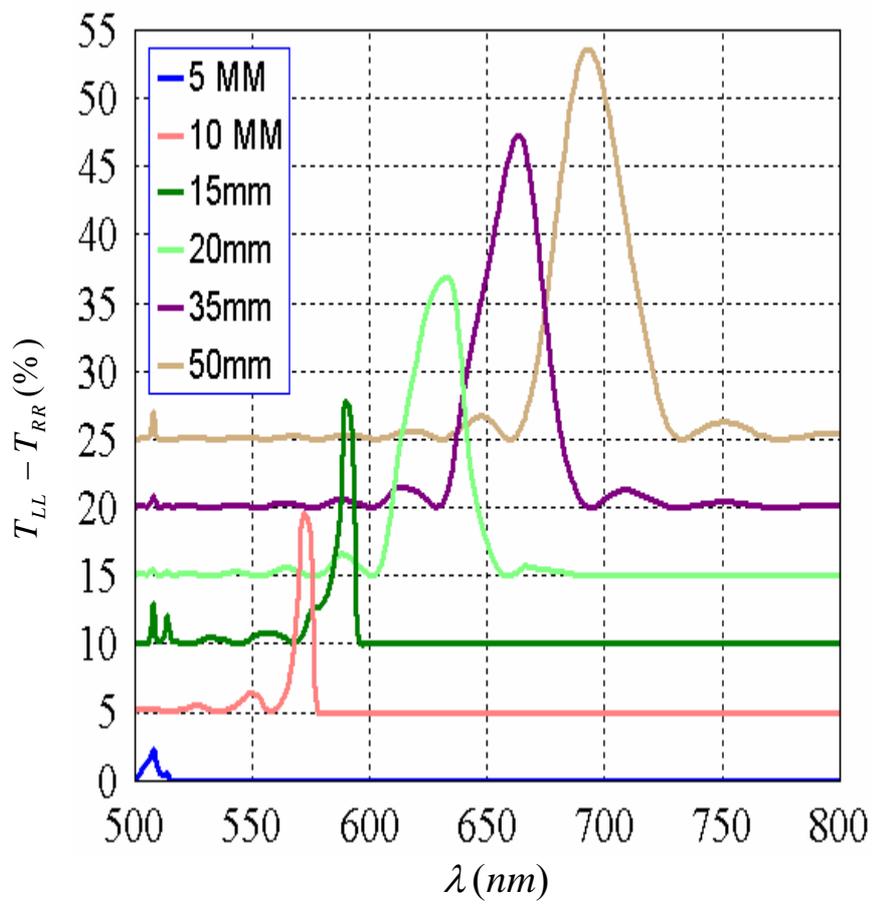

**Fig. 4. Babaei et al.**

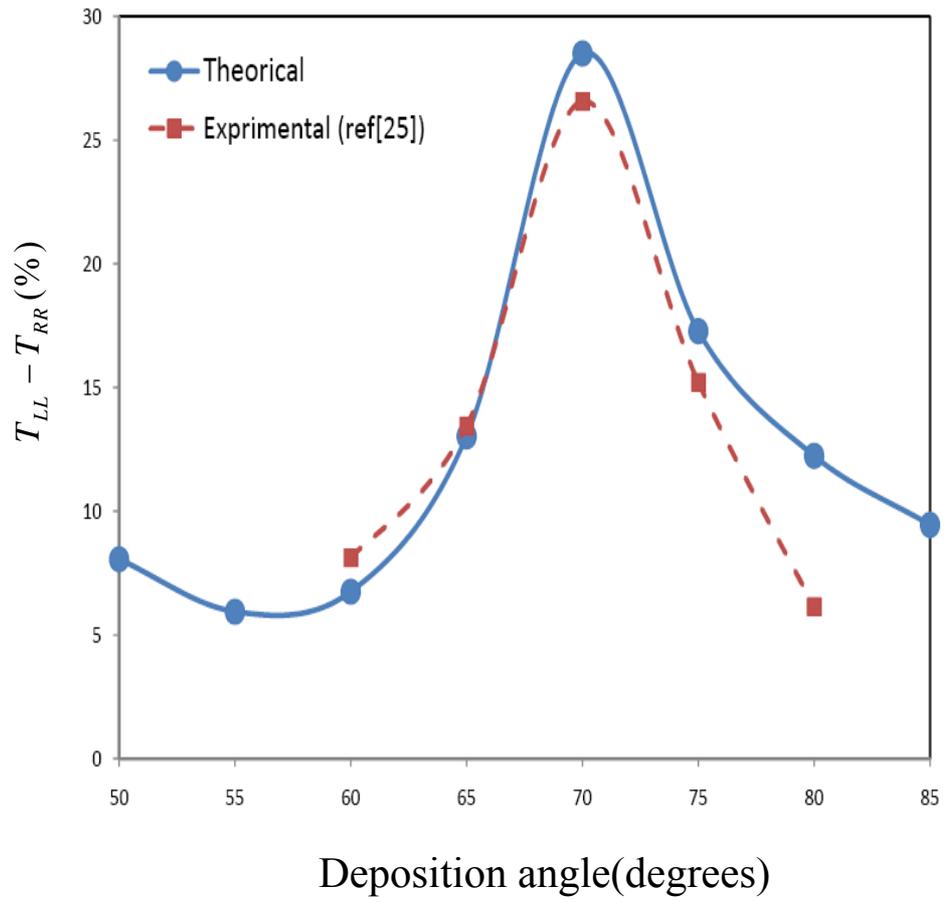

**Fig. 5. Babaei et al.**

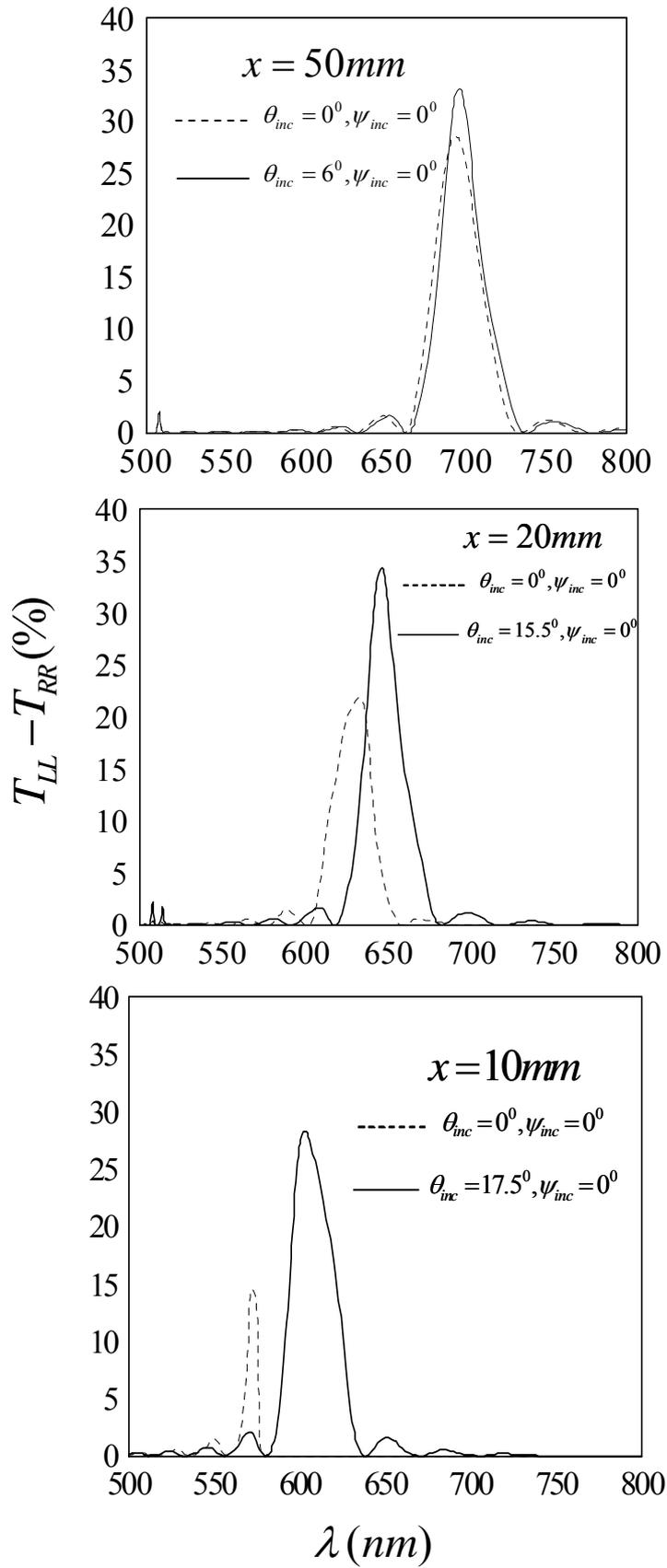

Fig. 6. Babaei et al.

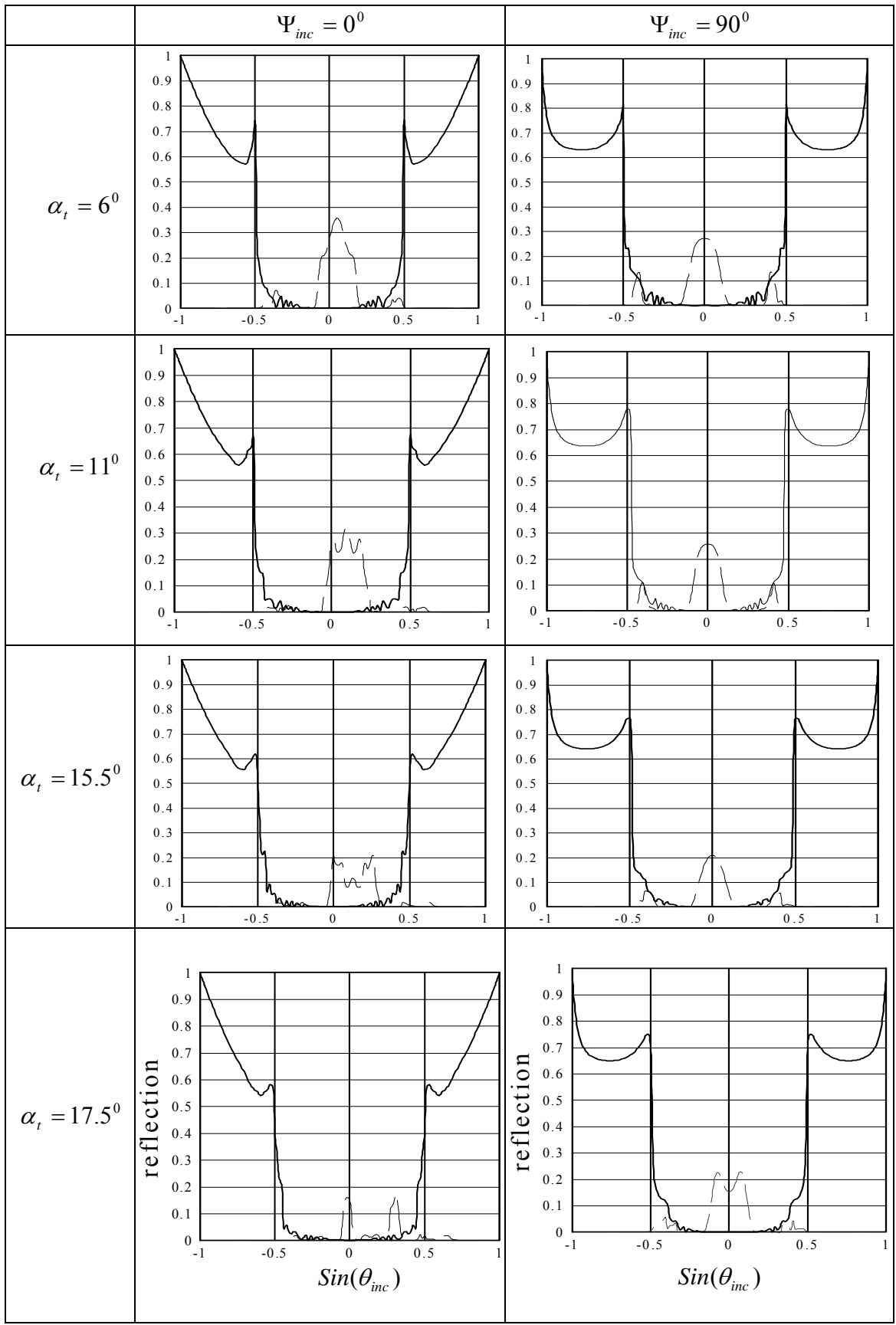

Fig. 7. Babaei et al.

**Solid line should be stronger. Lable "reflection" on all columns. Only left column is enough.**

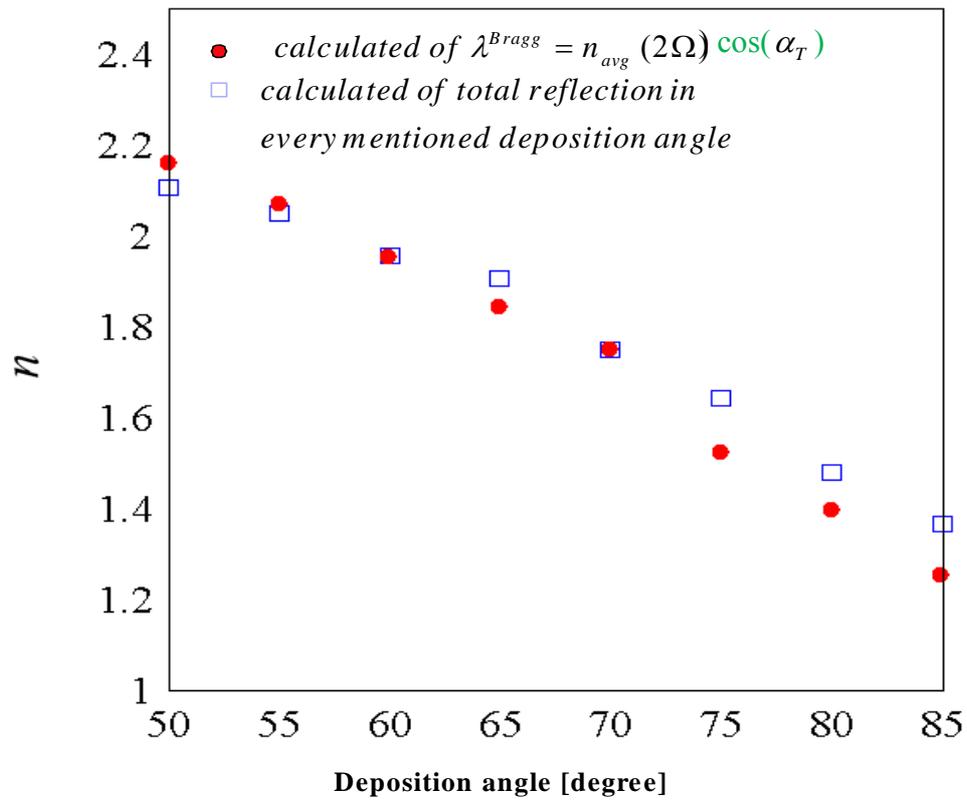

Fig. 8. Babaei et al.